**F-ion bridged Double-Decker Dysprosium Metallacrown with high-performance Single Molecule Magnet properties**

Yun-Xia Qu, Jin Wang, Ze-Yu Ruan, Guo-Zhang Huang, Yan-Cong Chen,, Jun-Liang Liu and Ming-Liang Tong*

___________________________________________________________________________

Key Laboratory of Bioinorganic and Synthetic Chemistry of Ministry of Education, School of Chemistry, Sun Yat-Sen University, 510006 Guangzhou, Guangdong, P. R. China.

*Corresponding authors. Email: tongml@mail.sysu.edu.cn.

**Abstract**

We report here a linear fluoride-bridged Double-Decker Dysprosium metallacrown with high-performance SMM. The successful introduction of stronger magnetic exchange-coupling in the axial direction, which is collinear with the Ising-type magnetic anisotropy axis of dysprosium ions, plays a pivotal role in improving the SMM properties of the double-decker Dysprosium metallacrown.

**Keywords:** single-molecule magnets (SMMs); metallacrown; Dysprosium; anisotropy; magnetic dipole

## 1. Introduction

Single-molecule magnets (SMM) have attracted enormous and extensive research interest from multiple disciplines[1-15]. Compared with conventional permanent bulk magnets, the magnetic properties of single-molecule magnets come from a single independent molecule, and there is no magnetic long-range order[16]. They not only exhibit superparamagnetism but also have the characteristics of monodispersity. Single-molecule magnets have higher magnetic storage density, are easier to chemically modify to achieve deviceization, and have special quantum effects[17]. It is obvious from the previous statements that the single-molecule magnets have potential application value and broad application prospects in the fields of high-density information storage, quantum computing, spin electronics and quantum bits[1-8].

Obviously, an excellent single-molecule magnet theoretically should have a longer magnetic relaxation time, at the same time a large and tunable effective energy barrier and high blocking temperature. Lanthanide-based single-molecule magnets (Ln-SMMs) have especially sparked increasing interest recently, due to their large magnetic moments and magnetic anisotropy[18-23]. It is firmly believed that the upper limit and ceiling of the relaxation time are restricted by the fast spin-lattice relaxation and quantum tunnelling of magnetization (QTM), especially for single-ion magnets that containing only one paramagnetic ion[13-14]. Then, specific symmetry elements are analysed for the elimination of transverse crystal fields and QTM, which are extremely effective strategy to improve the performance of single-molecule magnets[14]. Among them, single-molecule magnets with pentagonal and biconical geometry have shown remarkable achievements[8,24-27]. Inspired by these exciting results, our group has constructed a series of $D_{5h}$ SMMs based on the metallacrown (MC)[28-30]. Taking advantage of metallacrown approach as well as axially anisotropic ligation these SMMs show marvelous magnetic properties and behaviors.

In this report, we present the first double-decker Dysprosium metallacrown [Dy$_2$Cu$_{10}$(mquinha)$_{10}$(μ-F)F$_2$(py)$_{10}$](CF$_3$SO$_3$)$_3$ 0.5py 0.5CH$_3$OH 3H$_2$O (**1**, H$_2$mquinha = N-hydroxy-4-methylquinoline-2-carboxamide[31], py = pyridine) which introduces of strong ferromagnetic exchange in the direction of anisotropic axis. It hints at a promising and effective way to suppress QTM and improve SMMs properties by the rational choice of the complex symmetries: $D_{5h}$, in particular by avoiding crystal-field terms. On the other hand, the strong magnetic exchange between d-d and d-f metal ions also plays an indelible role in regulating and controlling the path of magnetic relaxation. In view of the observed anisotropic exchange interactions, it is expected that a more linear F-Dy-F geometry will lead to an improvement in the linear arrangement of the local magnetic anisotropy axes of F and [15-MC$_{Cu}$-5], which should have a strong impact on the strength and sign of the Dy-F anisotropic exchange couplings. In addition, in order to evaluate the contribution of the dipole/hyperfine coupling toward zero-field QTM, the yttrium diamagnetic diluted analogue {Dy$_{0.1}$Y$_{1.9}$Cu$_{10}$} (**1@Y**) were investigated by isothermal magnetization as well as time-dependent decay of magnetization measurements.

## 2. Results and Discussion

2.1 Structural analysis

Single-crystal and powder X-ray diffraction shows that title compounds are isostructural and crystallize in the monoclinic $P2_1/n$ space group (Table S1; Figures S1 and S2). The tetradentate hydroxamic acid ligand N-hydroxy-4-methylquinoline-2-carboxamide get rid of two protons and form mquinha$^{2-}$. It adopts the coordination mode of [32111][32] to chelate adjacent Cu$^{II}$ with two N and two O respectively (Figure 1d). Five such subunits self-assemble in a head-to-tail manner to form a neutral [15-MC$_{Cu}$-5] equatorial ring. Thanks to this coordination geometry, Dy$^{III}$ ions are encapsulated properly in a five-pointed star cavity provided by five hydroxyl oxygen atoms thus form a Dy[15-MC$_{Cu}$-5] unit. Meanwhile, 10 py ligands, located on both sides of MC ring and coordinated with Cu$^{II}$. A F$^-$ is linked with two units with 179.046° of Dy1-F-Dy2 and the other F$^-$ ions axially cap each side of the double-decker {Dy[15-MC$_{Cu}$-5]}$_2$ complex, which helps the stabilization of the pseudo-$D_{5h}$ symmetry of the first coordination sphere (Figure 1a and c). The value of continuous shape measures calculations (CShM)$^{33-34}$ is 0.506 and 0.498 for Dy1 and Dy2, respectively (Table S1). The average Dy–O distance is 2.409 and 2.4032 Å for Dy1 and Dy2, respectively. In addition, the Dy$^{III}$ ion is axially coordinated by two F$^-$ with a much shorter Dy–F length of 2.099 and 2.089 Å. The equatorial O–Dy–O angles are in the range of 71.86° to 72.00°, while the axial one is 174.74° to 175.22°, very nearly defining the 4f ion as a pentagonal bipyramid. All the Cu$^{II}$ ions are 6-coordinate with a [N4O2] octahedral geometry and the distances between the neighboring Cu$^{II}$ sites vary within 4.560–4.619 Å. Between the discrete molecules, the nearest Dy–Dy distance in the crystal structure is as far as 14.69 Å and there are no indirect or super-exchange interactions through intramolecular chemical bonds.

CCDC 2083697 (**1**) and 2083699 (**1@Y**) contain the supplementary crystallographic data for this paper. These data can be obtained free of charge from the Cambridge Crystallographic Data Centre *via* https://www.ccdc.cam.ac.uk/structures/.

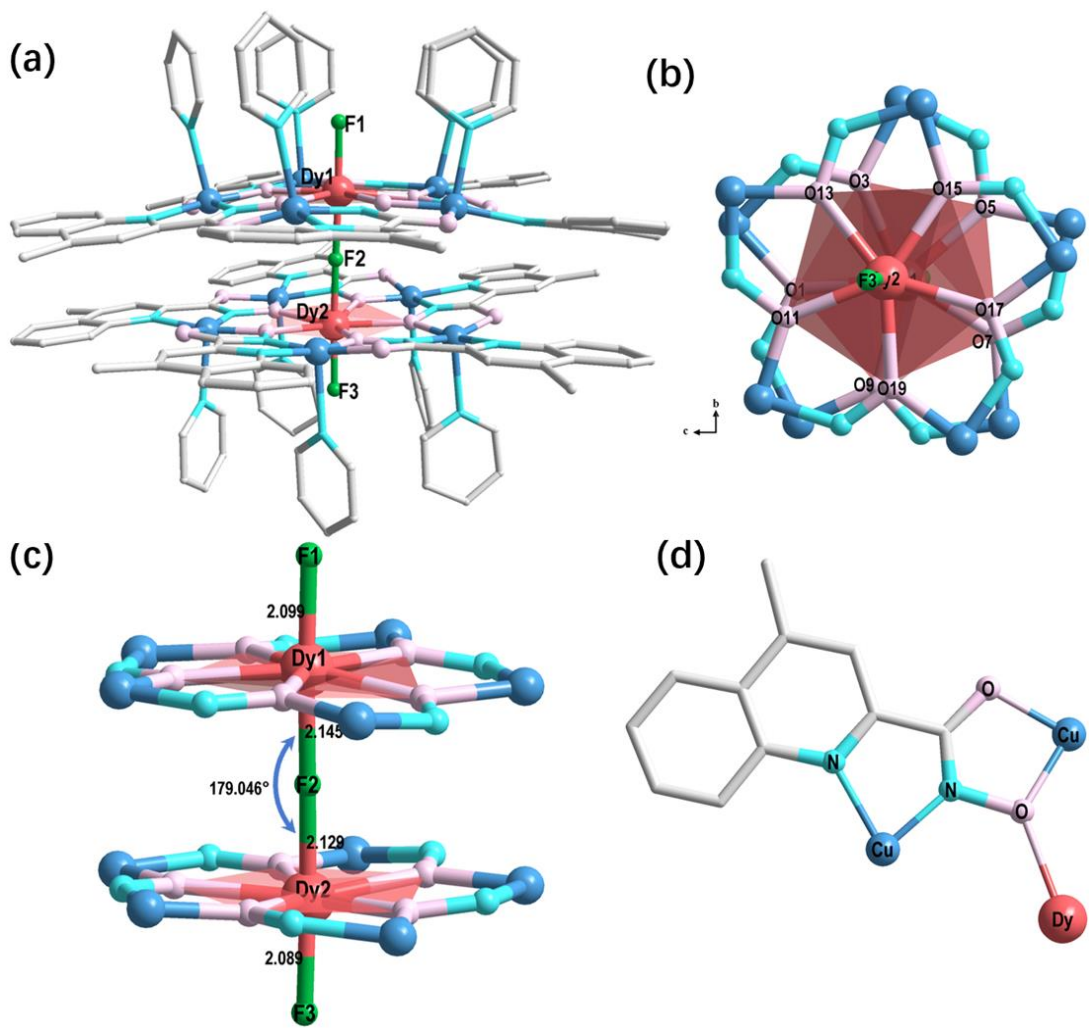

**Figure 1.** (a). The molecular structure of **1**. (b)Top and (c) side view of [15-MC$_{Cu}$-5] double-decker metallacrown core. (d). The [3.2111] coordination mode of mquinha$^{2-}$ ligand. The anion, solvent molecules and hydrogen atoms are omitted for clarity. Color code: Dy, red; O, pink; Cu, dark blue N, light blue; F, green; C, grey.

2.2 Magnetic properties

Temperature-dependent direct-current (DC) susceptibilities of both **1** and **1@Y** were performed. At room temperature the $\chi_M T$ products for **1**, ca. 32.14 cm$^3$ K mol$^{-1}$(Figure. 2), which is exactly in line with the expected value (the sum of two dysprosium and ten copper: $^6H_{15/2}$, $g_J = 4/3$ state free Dy$^{III}$ 14.17 cm$^3$ K mol$^{-1}$ and Cu$^{II}$ 0.38 cm$^3$ K mol$^{-1}$). On cooling, $\chi_M T$ for **1** show a gradual decline owing to a combination of the crystal field splitting of the $^6H_{15/2}$ ground multiplet of Dy$^{III}$ and the antiferromagnetic interactions among the Cu$^{II}$ ions. There is a clear rise in $\chi_M T$ below 50 K for **1**, indicating the presence of strong ferromagnetic also sundry and complex exchange interactions between metal ions. The dense structure of the coordination complex makes the Dy-F distance shorter, and the double-layer structure increases the number of bridging atoms, resulting in stronger magnetic exchange. It reaches a maximum of 37.85 cm$^3$ K mol$^{-1}$ at the lowest temperature of 1.8K. At low temperature, the magnetization of compound gradually increases on ramping magnetic field, with the maximum of 12.88 $N\mu_B$ at 5 K (Figure. 2 inset).

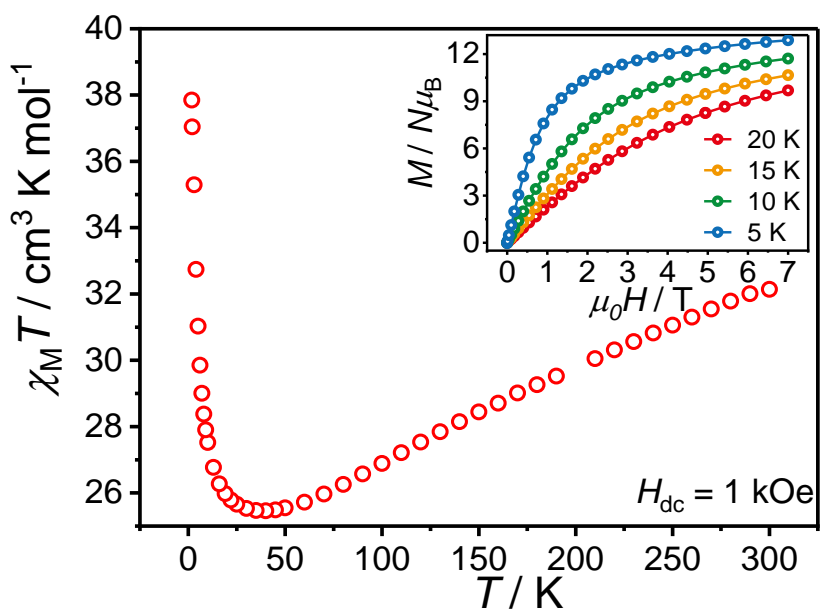

**Figure 2.** Plot of $\chi_M T$ vs $T$ under $H_{dc} = 1000$ Oe (main) and $M$ vs $H$ (inset) for **1** at various temperatures.

Because of long relaxion time at low temperature, magnetization hysteresis loops-the important phenomenon characteristic of magnetic bistability of magnets, were observed below 10 K on powder sample **1** and **1@Y** under a scan rate of 0.02 T s$^{-1}$ (Figure. 3). The hysteresis loops of samples **1** (Figure 3a) is obvious open at zero field below 6 K (with the cut-off criterion of $Hc = 0.01$ T). With the increase of temperature, the opening of hysteresis loop gradually decreases, which is manifested by the decrease of remanent magnetization ($M_r$) and coercive field ($H_c$). As for Y(III)-doped samples it exhibits butterfly-shaped hysteresis loops with close above 4K. Because of fast QTM, the hysteresis loops show no remanence at zero-field.

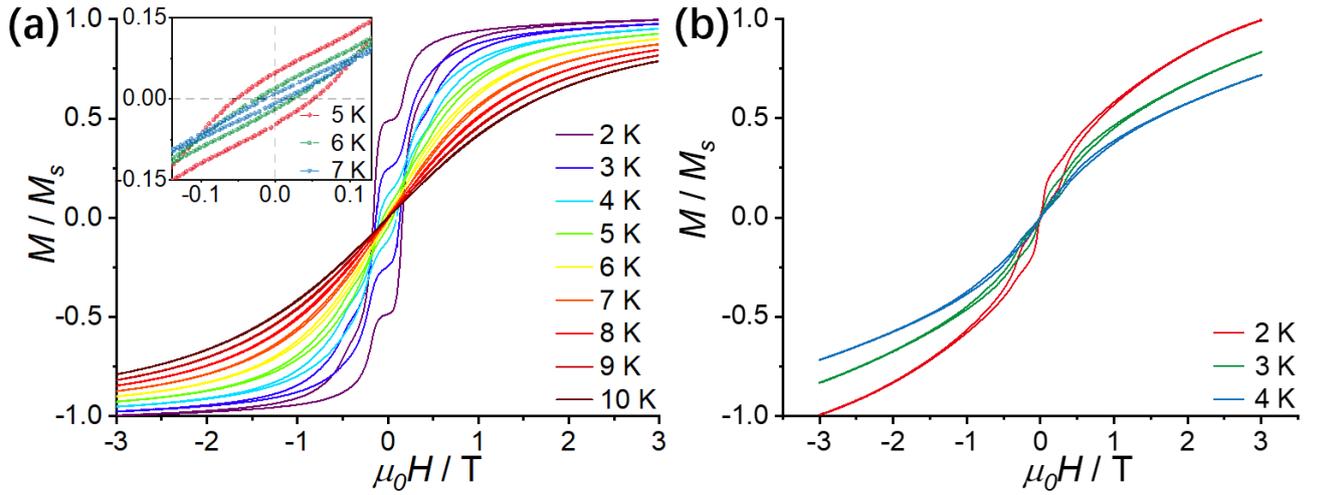

**Figure 3.** Hysteresis loops recorded on powder sample for **1** (a) and **1@Y** (b) at a sweep rate of 0.02 T s$^{-1}$ and the indicated temperatures. Inset: the enlargement of magnetic hysteresis loops.

To get further insight into magnetization dynamics, the alternative-current (ac) and direct-current (dc) magnetic susceptibility measurements were carried out on **1** and **1@Y** ranging 0.1-1488 Hz. From the temperature-dependent data (Figure 4a), the patterns are quite typical for high-performance SMMs, with the 1488 Hz peaking at 64 K and the 0.1 Hz peaking at 5 K under 0 dc filed. And there are no "tails" observed on the ac data in the measured window, suggesting zero-field QTM is indeed significantly precluded. For the frequency-dependent measurement, the ac peaks still keep slow shifting towards the low frequency region (see Supporting Information for details). The ln($\tau$) plots deviate from linearity at lower temperature suggests the presence of under-barrier processes. We can fit the magnetization relaxation with multiple relaxation mechanisms: two Orbach and a Raman process ($\tau^{-1} = \tau_{0(1)}^{-1}\exp(-U_{eff(1)}/k_BT) + \tau_{0(2)}^{-1}\exp(-U_{eff(2)}/k_BT) + CT^n$) to give $\tau_{0(1)} = 6.46(4) \times 10^{-3}$ s, $U_{eff(1)}/k_B = 29.4(4)$ K, $\tau_{0(2)} = 1.2(4) \times 10^{-11}$ s, $U_{eff(2)}/k_B = 1027(22)$ K, $C_{0Oe} = 4.57(5) \times 10^{-5}$ s$^{-1}$ K$^{-n}$, $n_{0Oe} = 4.0(3)$ (Figure. 4b). Meanwhile at high temperatures the relaxation time obeys the Arrhenius law, giving the best fit results of $\tau_0 = 2.8(1) \times 10^{-11}$ s and $U_{eff}/k_B = 967(25)$ K for zero dc field (Fig. 3c). Further investigations toward magnetic relaxation mechanisms of **1** and **1@Y** were obtained through *ab initio* calculations (see

Supporting Information for details).

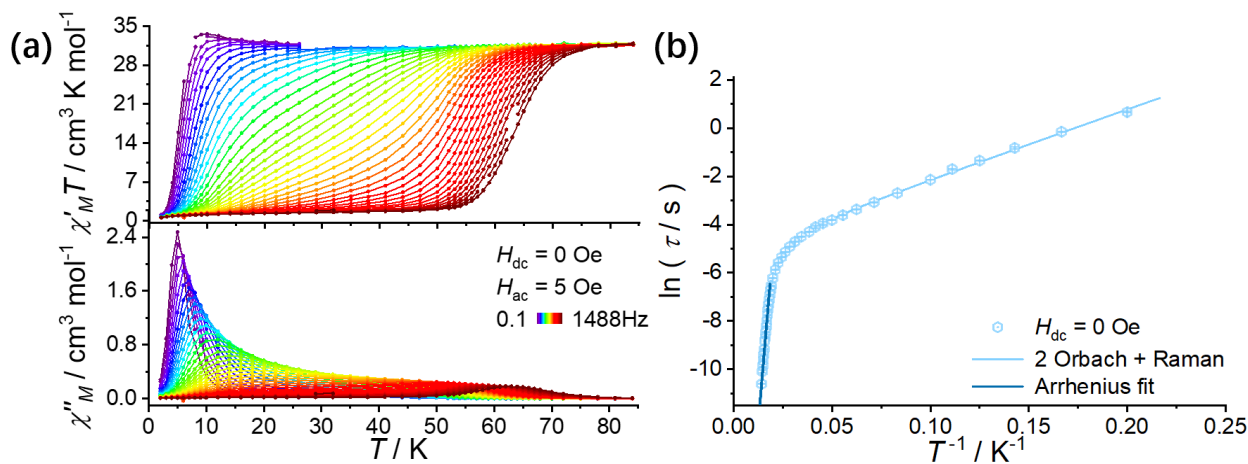

**Figure 4.** (a). Plots of the in-phase ($\chi_M'T$) product and out-of-phase ($\chi_M''$) susceptibility versus temperature under frequency of 0.1-1488 Hz at $H_{dc}$ = 0 Oe ($H_{ac}$ =5 Oe) for **1**. (b). Temperature dependence of the relaxation time $\tau$ (logarithmic scale) versus $T^{-1}$ in zero dc field (blue), the light blue solid lines are the best fit to the multiple relaxation equation, and dark blue one is the best fit to Arrhenius law.

## 3. Conclusions

In summary, we have constructed a linear F⁻ bridged 3d-4f double-decker metallacrown single molecule magnet by using the strategy of partition ligand method. The Dy(III) ion is chelated with the cavities of the metallacrown ring on the plane, and coordinated with the F⁻ ion end group in the axial direction, forming a flattened pentagonal bipyramid coordination geometry. The magnetic test shows that the QTM of Kramer ion Dy is effectively suppressed. On the one hand, metallacrown have great potential and advantages in the construction of excellent single molecule magnets. On the other hand, the strong F-F exchange in the axial direction greatly enhances the anisotropy.

**Conflicts of interest**

There are no conflicts to declare.

**Acknowledgements**

This work was supported by the National Key Research and Development Program of China (2018YFA0306001), the NSFC (Grant no. 21620102002 and 21822508), the Pearl River Talent

**Author contributions**

Ming-Liang Tong conceived the research and designed the project. Yun-Xia Qu carried out the synthesis and other experiments. Ze-Yu Ruan, Guo-Zhang Huang and Jun-Liang Liu contributed to magnetic measurements and analysis. Yun-Xia Qu wrote the manuscript with revisions and comments from the other authors.

# Synopsis

**F-ion bridged Double-Decker Dysprosium Metallacrown with high-performance Single Molecule Magnet properties**

Yun-Xia Qu, Jin Wang, Ze-Yu Ruan, Guo-Zhang Huang, Yan-Cong Chen,, Jun-Liang Liu* and Ming-Liang Tong*

First double-decker Dysprosium metallacrown which introduces of strong ferromagnetic exchange in the direction of anisotropic axis by F-.

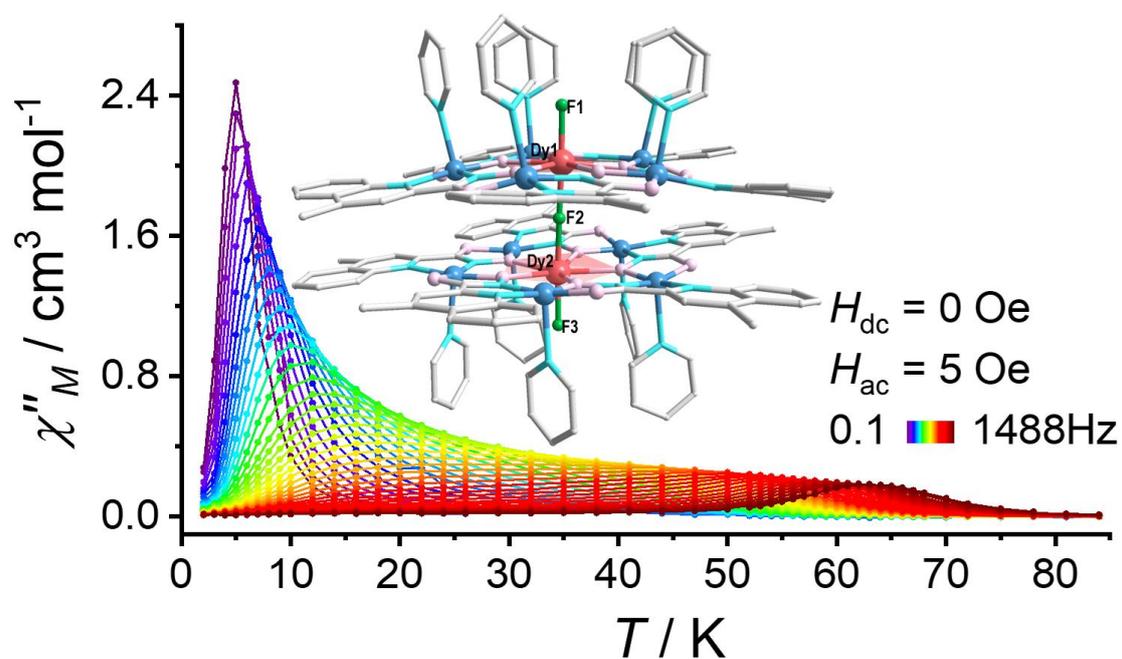